# Momentum-Resolved Sum-Frequency Vibrational Spectroscopy of Bonded Interface Layer at Charged Water Interfaces

Yao Hsiao[†], Ting-Han Chou[†‡], Animesh Patra, and Yu-Chieh Wen[*]

Institute of Physics, Academia Sinica, Taipei 11529, Taiwan, R. O. C.

*Correspondence to: ycwen@phys.sinica.edu.tw (Y.C.W.)

[†]Equal contribution

[‡]Present Address: Max Planck Institute for the Structure and Dynamics of Matter, 22761 Hamburg, Germany.

## ABSTRACT

Interface-specific hydrogen- (H-)bonding network of water next to a substrate (including air) directly controls the energy transfer and chemical reaction pathway at many charged aqueous interfaces. Yet, experimental characterization of such bonded water layer structure is still a challenge due to the presence of the ion diffuse layer. We now develop a sum-frequency (SF) spectroscopic scheme with varying photon momentums as an all–optic solution for retrieving the vibrational spectra of the bonded water layer and the diffuse layer, and hence microscopic structural and charging information about an interface. Application of the method to a charged surfactant–water interface reveals a hidden weakly-donor-H-bonded water species, suggesting an asymmetric hydration-shell structure of fully solvated surfactant headgroups. In another application to a zwitterionic phosphatidylcholine (PC) lipid monolayer–water interface, we find a highly polarized bonded water layer structure associating to the PC headgroup, while the diffuse layer contribution is experimentally proven to be negligible. Our all-optic method offers not only an *in situ* microscopic probe of the electrochemical and biological interfaces, but also a new opportunity for promoting these researches toward high spatial and temporal resolutions.

**KEYWORDS**
Sum-frequency vibrational spectroscopy, water interface, bonded interface layer

**TEXT**

Charged water interfaces are responsible for many natural phenomena[1,2] and of great importance in the development of advanced catalysts and energy storage devices.[3-5] Microscopic structure of such interfaces is in general under the influences of two effects: One is the interface-specific bonding interactions among water molecules, solvated ions, and the substrate effectively within a few monolayers of water away from the interface. This region is labelled as "the bonded interface layer (BIL)".[6] Another effect is the dc electric field, $E_0$, set up by interfacial charges and diffuse ions in a much extended region [so-called "diffuse layer (DL)"].[5] It has been recognized that the BIL structure controls the microscopic pathways and equilibrium of interfacial reactions and, therefore, dictates diverse chemical and biological processes.[1-5,7-10] For example, photoionization reaction of phenol proceeds $10^4$ times faster at the water surface than in the bulk aqueous phase owing to distinct hydration environments.[9] Water molecules at biomembranes assists the in-plane proton conductivity along the membrane that is vital for cellular bioenergetics.[2,10] Despite its importance, current knowledge on microscopic structure and the interplay between ionic and molecular species in the BIL is very limited due to paucity of nowadays experimental techniques.[8,11-13] In particular, scanning probe microscopies[11] and X-ray adsorption spectroscopy[12] were employed to probe the local surface structure and populations of donor-H-bond water species at solid surfaces, respectively, but the results do not yield information on the geometry or strength of water H bonds. On the other hand, infrared and Raman spectroscopies cannot distinguish the BIL from the DL.[13]

Vibrational sum frequency generation (SFG) has been developed as a versatile surface analytical tool and adopted to study the BIL at various (charge-neutral) aqueous surfaces.[9,14] At charged water interfaces, however, the surface specificity of this method is largely degraded because emergence of the dc surface field in the DL enables SFG from bulk water via the third-order optical nonlinearity, $\overleftrightarrow{\chi}^{(3)}$. Interference between the SF fields from BIL and DL not only hides vibrational signatures of the BIL[6,15] but also causes artificial resonances in the interfered spectra.[16] It appears crucial to separate the two SF contributions from a charged water interface, while it remains difficult so far.[6,15-22] In many reports, one of the two contributions was simply assumed constant or negligible.[17-19] In others, the DL was intentionally screened by adding considerable (external) ions in solutions, but the consequent disturbances of the BIL structure and chemistry therein were hardly addressed.[20,21] In 2016, Tian and Shen group proposed a SFG scheme for retrieving the (intrinsic) BIL spectra, which requires *prior* information about the surface charge density.[6] The idea was first demonstrated using carboxylic-acid monolayer on water[6], and later realized by Gibbs group for the silica/water interface with helps from the electrokinetic measurements.[15] A different gate-control scheme was recently proposed by Liu group using an electrolyte–insulator–semiconductor field-effect transistor (FET), which allows removal of the DL as the BIL structure kept effectively the same.[22] Despite their success, the demonstrations in Ref. [6] and [22] relied on specific interface systems (that allow spectroscopic quantification of all charged surface species or integration with a FET). While combination of SFG and electrokinetic measurements is ideal for better applicability[15], comparing the two independent measurements would be difficult as knowing that the local surface structure could be easily influenced by surface preparations[23] and measurement conditions[24,25], and may vary sample-by-sample.[26]

In this paper, we develop an all-optic scheme for direct and *in situ* determination of the vibrational spectra of BIL at various water interfaces. The scheme, labeled as "momentum-resolved SF vibrational spectroscopy (MR-SFVS)", is realized by detecting a set of SF radiations from a sample with a varying input photon momentum. This idea, first proposed in 2016[6], was recently examined theoretically in a thorough manner.[27] Here we experimentally prove its validity and demonstrated its viability to different interfaces. Its application to zwitterionic phosphatidylcholine (PC) lipid/water interfaces shows a highly polarized bonded water layer structure associating to the PC headgroup, while the DL contribution to SFG is proven to be negligible. In another application to a charged surfactant–water interface, the result reveals a hidden weakly-donor-H-bonded water species in the BIL, revealing the hydration structure of the surfactant headgroups. This technique offers a viable means for *in situ* probing of electrochemical and aqueous biological interfaces at the molecular level.

**Theory**

We follow Ref. [6] to formulate SFG process at a charged water interface next to an isotropic medium with $z$ along the surface normal. The reflected SFG from such an interface has its field proportional to the effective surface nonlinear susceptibility, $\overleftrightarrow{\chi}^{(2)}_{S,eff}(\omega_{SF} = \omega_1 + \omega_2)$, which can be expressed as

$$\overleftrightarrow{\chi}^{(2)}_{S,eff} = \overleftrightarrow{\chi}^{(2)}_{BIL} + \int_{0^+}^{\infty} \overleftrightarrow{\chi}^{(3)}(z') \cdot \hat{z} E_0(z') e^{i\Delta k_z z'} dz' = \overleftrightarrow{\chi}^{(2)}_{BIL} + \overleftrightarrow{\chi}^{(2)}_{DL}, \quad (1)$$

$$\overleftrightarrow{\chi}^{(2)}_{DL} \equiv \int_{0^+}^{\infty} \overleftrightarrow{\chi}^{(3)}(z') \cdot \hat{z} E_0(z') e^{i\Delta k_z z'} dz' \cong \overleftrightarrow{\chi}^{(3)}_B \cdot \hat{z} \Psi,$$

with $\Psi \equiv \int_{0^+}^{\infty} E_0(z') e^{i\Delta k_z z'} dz'$.

Here $\overleftrightarrow{\chi}^{(2)}_{BIL}$ and $\overleftrightarrow{\chi}^{(2)}_{DL}$ denote the nonlinear optical susceptibilities of the BIL (at $z = 0^- \sim 0^+$) and DL, respectively. $\Delta k_z = k_{SF,z} + k_{1,z} + k_{2,z}$ describes the phase mismatch for reflected SFG. We have considered (*i*) the contribution of the DL (approximately, at $z > 0^+$) is characterized by the third-order nonlinear susceptibility of bulk water[28], $\overleftrightarrow{\chi}_B^{(3)}$, and (ii) $\left|\overleftrightarrow{\chi}_{BIL}^{(2)}\right| \gg \left|\int_{0^-}^{0^+} \overleftrightarrow{\chi}^{(3)}(z') \cdot \hat{z} E_0(z') dz'\right|$ in the BIL due to stronger influences of the H bonding compared to the mean-field-induced electrostatic energy.[15]

To separate $\overleftrightarrow{\chi}_{BIL}^{(2)}$ and $\overleftrightarrow{\chi}_{DL}^{(2)}$ in Eq. (1), we compare two measurements with different $\Delta k_z$ (labeled below for all symbols with a superscript *A* or *B*), which are related through Eq. (1) by

$$\overleftrightarrow{\chi}_{S,eff}^{(2),B} = \overleftrightarrow{\chi}_{S,eff}^{(2),A} + \Delta\overleftrightarrow{\chi}_{DL}^{(2)}, \qquad (2)$$

$$\Delta\overleftrightarrow{\chi}_{DL}^{(2)} \equiv \overleftrightarrow{\chi}_{DL}^{(2),B} - \overleftrightarrow{\chi}_{DL}^{(2),A} = \overleftrightarrow{\chi}_B^{(3)} \cdot \hat{z} \int_{0^+}^{\infty} E_0(z') \left(e^{i\Delta k_z^B z'} - e^{i\Delta k_z^A z'}\right) dz'.$$

It is clear that $\Delta\overleftrightarrow{\chi}_{DL}^{(2)}$ senses $E_0(z)$ with a weighting factor $\left(e^{i\Delta k_z^B z} - e^{i\Delta k_z^A z}\right)$ that approaches to zero as $z \ll \Delta k_z^{-1}$ (= coherence length $l_c$). It indicates that $\Delta\overleftrightarrow{\chi}_{DL}^{(2)}$ cannot be sensitive to subtle ionic packing structure and potential profile at immediate neighborhood of the interface, as proposed by many electric-double-layer (EDL) models[5,29], but reflects essentially the Poisson-Boltzmann (PB) ion distribution on the length scale of $l_c$ (about tens of nanometers), i.e., the diffuse layer. Therefore, $\Delta\overleftrightarrow{\chi}_{DL}^{(2)}$ represents a differential contribution from the DL detected with two *effective* probing depths ($l_c^A$ versus $l_c^B$, as sketched in Fig. 1A). It offers a probe to the potential drop across the DL, $\phi_0$, and the net surface charge density with the non-PB-distributed interfacial ions included.

Without loss of validity, we propose to measure complex $\overleftrightarrow{\chi}_{S,eff}^{(2),A}$ and intensity $\left|\overleftrightarrow{\chi}_{S,eff}^{(2),B}\right|^2$ spectra for a given sample. With the previously measured $\overleftrightarrow{\chi}_B^{(3)}(\omega_2)$ of bulk water[6], we can use Eq. (2) to fit the measured $\left|\overleftrightarrow{\chi}_{S,eff}^{(2),B}(\omega_2)\right|^2$ spectrum with the complex $\overleftrightarrow{\chi}_{S,eff}^{(2),A}(\omega_2)$ through the PB

theory[5] that relates $E_0(z)$ to $\phi_0$ and the ionic strength. Having the only fitting variable $\phi_0$ deduced, we then follow Ref. [6] to estimate $\overleftrightarrow{\chi}_{DL}^{(2),A}(\omega_2)$ from Eq. (1) with the known $\overleftrightarrow{\chi}_B^{(3)}(\omega_2)$ and the PB theory, and subsequently determine the complex $\overleftrightarrow{\chi}_{BIL}^{(2)}(\omega_2)$ spectrum from the difference between the measured $\overleftrightarrow{\chi}_{S,eff}^{(2),A}(\omega_2)$ and the estimated $\overleftrightarrow{\chi}_{DL}^{(2),A}(\omega_2)$. Note that the retrieved $\overleftrightarrow{\chi}_{BIL}^{(2)}(\omega_2)$ is independent of the EDL models in terms of the non-PB-distributed interfacial ionic structure assumed. We also remark that Hore and Tyrode attempted to analyze SF intensity spectra with a set of $\Delta k_z$ without phase information. No separate spectral information on BIL or the DL could be retrieved.[30]

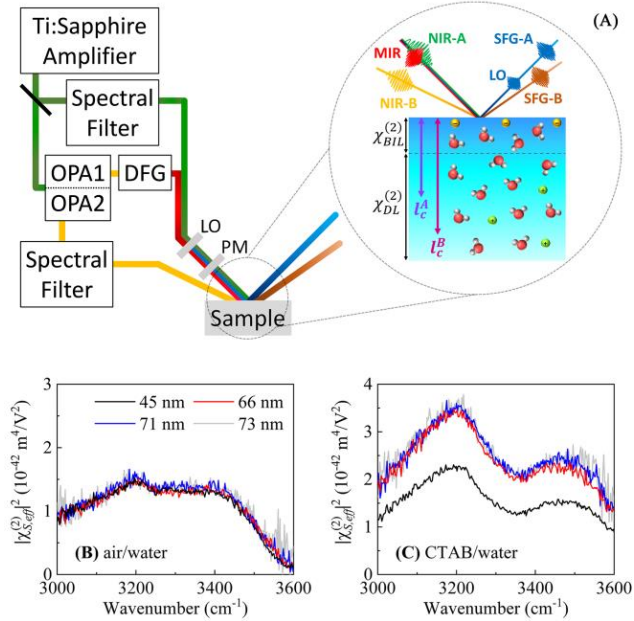

**Figure 1.** (*A*) Illustration of the MR-SFVS setup and a charged interfacial structure probed. The SF field, SFG-A (SFG-B), was emitted through mixing of a MIR pulse with one of the two near-infrared fields, NIR-A (NIR-B), and detected with heterodyne (homodyne) scheme with a coherence length of $l_c^A$ ($l_c^B$). LO and PM refer to local oscillator and phase modulator, respectively. The optical inputs were prepared by a Ti:sapphire amplifier equipped with two optical parametric amplifiers (OPAs), difference frequency generation (DFG), and two grating-based spectral filters. See "Methods" section for details. (*B*) and (*C*)

$\left|\chi_{S,eff}^{(2)}\right|^2$ SF intensity spectra of the surfaces of neat water and a 10 μM CTAB solution measured by MR-SFVS with different coherent lengths.

**Spectral decomposition by MR-SFVS experiment**

Our MR-SFVS experiments with $l_c^A$ of 45 nm and $l_c^B$ of 66 ~ 73 nm were performed *in situ* with a modified phase-sensitive SFVS setup with S-, S-, and P-polarized SF, near-infrared, and mid-infrared (MIR) fields. As sketched in Fig. 1A, the setup comprises two SFG configurations for different $\Delta k_z$. They share the same femtosecond MIR input but have separate narrow-band near-infrared input pulses with distinct wavelengths $\lambda$ and the angle of incidence $\theta$. (For $l_c^A$, $\lambda =$ 0.8 μm and $\theta =$ 45°; For $l_c^B$, $\lambda =$ 1.2~1.4 μm and $\theta =$ 64°. See "Methods" section for details.) The measured spectra of a sample were normalized against a *z*-cut quartz reference and shown below in MKS units after correcting the Fresnel coefficients. High spectroscopic specificity to $\Delta\chi_{DL}^{(2)}$ is first confirmed by inspecting the OH-stretch SF intensity spectra of the neat water/air interface versus a charged surfactant [cetyltrimethylammonium bromide (CTAB)]/water interface [see Supplementary Section 1 for sample preparation]. As shown in Fig. 1B, the measured spectra of the neat water surface are essentially the same upon changes in $\Delta k_z$, as expected for a charge-neutral interface with $\chi_{S,eff}^{(2),A} = \chi_{S,eff}^{(2),B} = \chi_{BIL}^{(2)}$. (See Supplementary Fig. S3 for raw data without Fresnel correction). By contrast, the charged CTAB/water interface shows a manifest correlation between $\Delta k_z$ and the SFG strength (Fig. 1C), i.e., the strength increases by ~40% (4 ~ 9%) with a notable (minor) increase in $l_c$ from 45 to 66 nm (from 66 to 73 nm). It is such a difference disclosing the emergence of a DL ($\chi_{S,eff}^{(2),B} = \chi_{S,eff}^{(2),A} + \Delta\chi_{DL}^{(2)}$) and being the basis for the spectral separation.

More quantitatively, we adopted a closely packed lignoceric acid ($C_{23}H_{47}COOH$) monolayer on water to verify the spectral separation by MR-SFVS. Deprotonation of this monolayer ($COOH \leftrightarrow COO^- + H^+$) with subphase pH and the resultant spectral changes have been well studied by SFVS[6,31]; It was shown that the interface is charge-neutral at low pH (< ~2), but increasingly negatively charged with pH upon deprotonation of the monolayer. Knowing that the fractional ionization of this monolayer is less than few percent at acidic pH (2~7)[6], the induced structural perturbation on the BIL is hardly detected. Therefore, we anticipate $\chi^{(2)}_{BIL}(\omega_2)$ retrieved from MR-SFVS in this pH region being invariant and similar as $\chi^{(2)}_{S,eff}(\omega_2) \cong \chi^{(2)}_{BIL}(\omega_2)$ at pH < 2, and expect the deduced $\phi_0$ versus pH to follow *pK$_a$* of the deprotonation reaction. Figure 2A shows the SF intensity spectra of the sample measured at three acidic pH in the OH-stretching region. They are very similar to the earlier report[6], but the effect of varying $\Delta k_z$ is presented here for the first time. It is seen that the nearly charge-neutral interface at pH 2.1, again, exhibits $\Delta k_z$-insensitive SF spectra, whereas the SF intensity at pH 4.1 and 5.7 varies prominently with $\Delta k_z$, revealing appearance of the DL due to the monolayer ionization.

We follow the two-step analysis described above to decompose the spectra. First, we use the measured $\chi^{(2),A}_{S,eff}(\omega_2)$ and the known[6] $\chi^{(3)}_B(\omega_2)$ to fit the measured $\left|\chi^{(2),B}_{S,eff}(\omega_2)\right|^2$ spectra via Eq. (2) with the PB theory[5] for deducing $\phi_0$. Secondly, we use Eq. (1) to estimate $\chi^{(2),A}_{DL}(\omega_2)$ from $\phi_0$ with the known $\chi^{(3)}_B(\omega_2)$ and the PB theory, and then obtain $\chi^{(2)}_{BIL}(\omega_2)$ from the difference between the measured $\chi^{(2),A}_{S,eff}(\omega_2)$ and the estimated $\chi^{(2),A}_{DL}(\omega_2)$. As seen in Fig. 2A, the fitting quality for the $\left|\chi^{(2),B}_{S,eff}\right|^2$ spectra of the two charged interfaces is reasonably well. (See Supplementary Fig. S4 for fits with different values of $\phi_0$.) The diffuse layer potential so obtained indicates that the interface is negatively charged with $\phi_0 = -70 \pm 7$ ($-191 \pm 11$) mV at pH 4.1 (5.7).

One could compare these estimates with the calculation via the deprotonation reaction equation for fatty acids with $pK_a$ of 5.1 ~ 5.6[6,31], which yields $\phi_0 = -77\pm10$ ($-170\pm10$) mV at pH 4.1 (5.7), and indeed find the agreement. (See Ref. [6] for the detailed calculation.) Furthermore, even with appreciable pH dependency of $\phi_0$ and the measured spectra (Fig. 2A), the complex $\chi_{BIL}^{(2)}$ vibrational spectra retrieved at pH 4.1 and 5.7 (Fig. 2B, 2C) are found consistent with $\chi_{BIL}^{(2)}$ ($\cong \chi_{S,eff}^{(2),A}$) at pH 2.1, as expected for a RCOOH–dictated H-bonding structure with trivial perturbations by sparse RCOO$^-$.

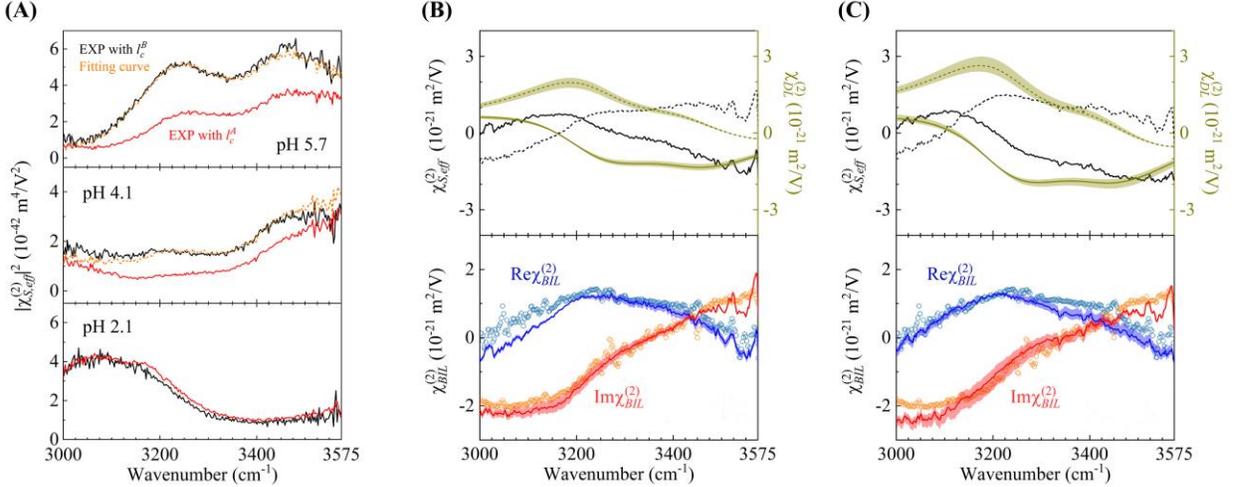

**Figure 2.** Analysis of the MR-SFVS results of the lignoceric acid monolayer/water interface. (*A*) $\left|\chi_{S,eff}^{(2)}\right|^2$ SF intensity spectra at pH 2.1, 4.1, and 5.7 measured with the coherence length of $l_c^A$ and $l_c^B$ (solid lines). Dashed lines denote the fits for $\left|\chi_{S,eff}^{(2),B}(\omega_2)\right|^2$ based on complex $\chi_{S,eff}^{(2),A}(\omega_2)$ from *B* and *C*. (*B*) and (*C*) The measured $\chi_{S,eff}^{(2),A}(\omega_2)$ and the retrieved $\chi_{DL}^{(2),A}(\omega_2)$ and $\chi_{BIL}^{(2)}(\omega_2)$ spectra at (*B*) pH 4.1 and (*C*) pH 5.7. Solid (dashed) lines in the upper panel denote real (imaginary) part of $\chi_{S,eff}^{(2),A}(\omega_2)$ and $\chi_{DL}^{(2),A}(\omega_2)$. In the bottom panel, the complex $\chi_{BIL}^{(2)}(\omega_2)$ retrieved at pH 4.1 or 5.7 (lines) is compared with $\chi_{BIL}^{(2)}(\omega_2)$ [$\cong \chi_{S,eff}^{(2)}(\omega_2)$] measured at pH 2.1 (dots). The shadowed regions in *B* and *C* denote uncertainty. Consistency of the retrieved $\chi_{BIL}^{(2)}(\omega_2)$ at the three acidic pH confirms validity of the MR-SFVS analysis.

**Application: Charge neutrality of zwitterionic PC lipid/water interfaces**

With validity of the spectral separation confirmed, one can now apply to the scheme to various aqueous interfaces without assumption or *prior* information about the interface. We offer two demonstrations and show that, even in well-investigated cases, the result could provide new information owing to the spectral separation. In the first case, we study charge neutrality of the zwitterionic PC lipid monolayers on water. Such an interface is known to have great biological relevance because of the key role of PC lipids in constituting the membranes. In many SFVS studies, the spectrum interpretation relied on nominal charge neutrality of the zwitterionic headgroup.[32,33] However, the zeta potential for liposomes of PC lipids has been reported to be negative at pH 7.[34] It was also argued that the PC monolayers on neat water were positively charged due to partial deprotonation of the phosphate groups with *pKa* in the range of 1 to 3.[35] Whether it is valid to presume negligible $\chi_{DL}^{(2)}(\omega_2)$ in interpreting the SF spectra appears to be questionable.

We revisit this issue by MR-SFVS qualitatively. To examine the intrinsic effects of the PC headgroup, the studied samples include two types of PC monolayers on water: 1,2-dipalmitoyl-*sn*-glycero-3-phosphocholine (DPPC) Langmuir monolayer and hexadecyl phosphocholine (HePC) Gibbs monolayer (see Supplementary Section 1 for the sample preparation). HePC has the same headgroup but differs from DPPC in terms of its single alkyl chain and lack of the glycerol backbone. Figure 3 shows the measured SF intensity spectra for the two types of monolayers on water at bulk pH ~7. Both interfaces are found to be invariant upon the change in $\Delta k_z$, evidencing an ignorable $\chi_{DL}^{(2)}(\omega_2)$ and, hence, the PC–water interfaces are essentially charge-neutral. Although this conclusion may not be very surprising, it is a prerequisite to accurate elucidation of the SF vibrational spectra.[32,33] For example, the HePC/water interface is seen to display very intense OH-stretching resonances ($3.4 \times 10^{-41}$ m$^4$/V$^2$ in Fig. 3), which is readily comparable to these from many charged aqueous interfaces with significant $\chi_B^{(3)}$ contributions (See Supplementary Fig. S1 for examples). Using MR-SFVS, one can now safely ignore the DL contribution and attribute the observed strong resonance *solely* to a highly polarized bonded water layer structure associating to the PC headgroup. This is likely due to the local dipolar field generated between the charged phosphate and choline and/or a H-bonding structure of water bridging both phosphate and choline, as pointed out by Morita, Bonn, and co-workers.[8,33,36] A brief discussion about the BIL structure via the $\text{Im}\chi_{BIL}^{(2)}$ ($\cong \text{Im}\chi_{S,eff}^{(2)}$) spectra is given in Supplementary Section 2.

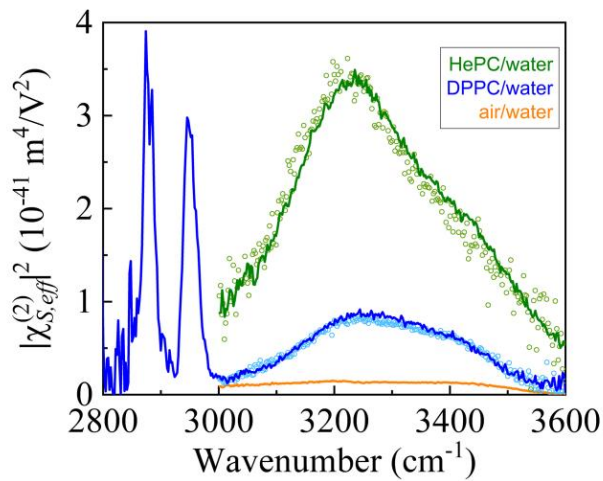

**Figure 3.** SF intensity spectra in the OH- and CH-stretching regions for a closely packed DPPC Langmuir monolayer/water interface and the HePC Gibbs monolayer on a 1.6 μM HePC solution with subphase pH of 7.0 and 7.8, respectively, in comparison with the neat water/air interface. Lines and dots denote the results measured with $l_c = l_c^A$ and $l_c^B$, respectively, and their consistency indicates a negligible DL and the responsible role of BIL in SFG. CH-stretching vibrations reveal nearly vertical alignment of the alkyl chains and negligible gauche defects in the DPPC monolayer.

**Application: BIL structure of surfactant/water interfaces**

In the second case, we revisit charged surfactant/water interfaces with a focus on the microscopic hydration structure of the adsorbed surfactants. This fundamental issue remains poorly understood on the molecular level[37,38] and is addressed here through $\text{Im}\chi_{BIL}^{(2)}(\omega_2)$ for the first time. (Note that it cannot be retrieved using zeta potential[15] that characterizes a dynamic system in the electrokinetic measurement where the surfactant may never reach adsorption equilibrium.[25])

We performed MR-SFVS measurements on the air/water interface with a series of CTAB concentrations in water and analyzed the results via the two-step analysis described above to deduce $\phi_0$, $\chi_{DL}^{(2),A}(\omega_2)$, and $\chi_{BIL}^{(2)}(\omega_2)$ in order. Shown in Fig. 4 and 5A are the measured $\left|\chi_{S,eff}^{(2)}\right|^2$ and $\text{Im}\chi_{S,eff}^{(2)}$ OH-stretching spectra, respectively, which are essentially similar to these reported earlier for given CTAB concentrations[39] (less than 3% of the critical micelle concentration). From Fig. 4, the difference of the SF intensity for $\Delta k_z = \Delta k_z^A$ versus $\Delta k_z^B$ increases with the bulk CTAB concentration, reflecting increasing $|\Delta\chi_{DL}^{(2)}|$ and surface charge density due to CTA$^+$ adsorption. We can, again, fit the $\left|\chi_{S,eff}^{(2),B}(\omega_2)\right|^2$ spectra reasonably well with the measured $\chi_{S,eff}^{(2),A}(\omega_2)$ and the known[6] $\chi_B^{(3)}(\omega_2)$, yielding increasingly positive-valued $\phi_0$ (see Supplementary Fig. S4B and S5). As for $\text{Im}\chi_{S,eff}^{(2),A}(\omega_2)$ in Fig. 5A, it exhibits an obvious bipolar spectral distortion with respect to the neat water surface upon surface adsorption of CTAB. By decomposing it into $\text{Im}\chi_{DL}^{(2),A}(\omega_2)$ and $\text{Im}\chi_{BIL}^{(2)}(\omega_2)$ with the deduced $\phi_0$, we verify that the observed change of $\text{Im}\chi_{S,eff}^{(2),A}(\omega_2)$ comes mainly from the emergent $\text{Im}\chi_{DL}^{(2),A}(\omega_2)$ (Fig. 5B), and the bipolar spectral shape of $\text{Im}\chi_{DL}^{(2),A}(\omega_2)$ is not real vibrational signatures but a consequence of absorptive-dispersive mixing of $\chi_B^{(3)}(\omega_2)$ via complex $\Psi$ through Eq. (1).[16] For example, a positive hump in $\text{Im}\chi_{DL}^{(2),A}(\omega_2)$ at ~3600 cm$^{-1}$ is an artificial resonance owing to the contribution from $\text{Re}\chi_B^{(3)}(\omega_2)$.

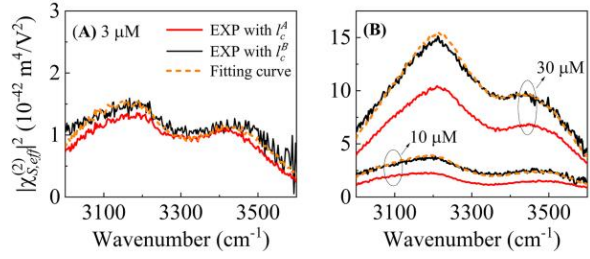

**Figure 4.** $\left|\chi_{S,eff}^{(2)}\right|^2$ SF intensity spectra of the surfaces of (*A*) 3 μM and (*B*) 10 and 30 μM CTAB solutions measured with $l_c = l_c^A$ and $l_c^B$ (solid lines). Dashed lines denote the fits for $\left|\chi_{S,eff}^{(2),B}(\omega_2)\right|^2$ based on complex $\chi_{S,eff}^{(2),A}(\omega_2)$ from Fig. 5A.

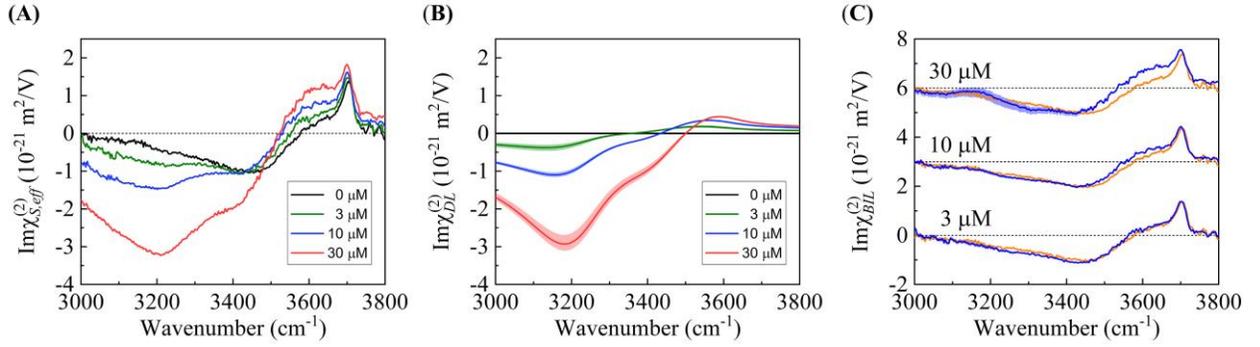

**Figure 5.** Analysis of the MR-SFVS results of the air/water interface with different CTAB concentrations in water. (*A*) The measured $\text{Im}\chi_{S,eff}^{(2),A}$ spectra and (*B*) the retrieved $\text{Im}\chi_{DL}^{(2),A}$ spectra (solid lines). (*C*) The $\text{Im}\chi_{BIL}^{(2)}$ spectra retrieved from *A* and *B* (blue lines), in comparison with that of the neat water surface (orange lines). The data are vertically shifted for clarity. The shadowed regions in *B* and *C* denote uncertainty, and the dashed lines in *A* and *C* depicts zero.

With $\text{Im}\chi_{DL}^{(2),A}(\omega_2)$ removed from $\text{Im}\chi_{S,eff}^{(2),A}(\omega_2)$, unambiguous vibrational signatures of the interface can be identified through the $\text{Im}\chi_{BIL}^{(2)}(\omega_2)$ spectra (Fig. 5C). They look similar to the neat water surface but display modified resonances above 3400 cm$^{-1}$ for [CTAB] $\geq$ 10 μM, disclosing surfactant-induced structural perturbation on the BIL. We find that this spectral modification is dominated by an emergent positive band at ~3610 cm$^{-1}$ with a linewidth of ~240 cm$^{-1}$, accompanying with tiny changes in the free OH resonance (that red-shifts by 4±2 cm$^{-1}$ with its amplitude reduced by 16±6 % for [CTAB] = 30 μM, as revealed by spectrum analyses based on approximation of discrete Lorentzian resonances. See Supplementary Section 3 for details).

The new positive 3610–cm$^{-1}$ band must result from OH stretching of weakly donor-bonded water molecules with their H atoms pointing upward (toward air). In interpreting its origin, one may consider water molecules associating to hydrophobic tail or headgroup of the surface CTAB. The former is likely irresponsible because the involved van der Waals interaction is usually too weak to cause obvious spectral changes more than perturbation of the dangling OH frequency.[18] As for the latter, we consider first an asymmetric hydration-shell structure of a fully solvated surfactant headgroup and then other possibilities. As sketched in Fig. 6, adsorption of CTA$^+$ to the water surface breaks the up-down symmetry of its hydration shell. The bottom part of the shell comprises water molecules with net downward-pointing OHs H-bonded to their water neighbors, whereas the upper counterpart of the shell is likely to include more H-up-oriented water molecules exposing to air (or the alkyl chain) and thus rarely (or weakly) donor-H-bonded. The distinct bonding conditions cause a relative frequency shift, so that the spectral contributions from the upper/bottom parts of the hydration shell are unlikely to fully compensate each other. The high-frequency residual dominated by the upper part of the hydration shell can thus serve an explanation for the new positive band at ~3610 cm$^{-1}$. The moderate red shift and broadening of this band with respect to the free-OH peak can be understood with association of the water molecules to the CTA$^+$ through the water oxygen taken into account.[40] On the other hand, this scenario may imply spectral features from the bottom part of the hydration shell at relatively lower frequencies, which is not clearly identified in the retrieved spectra. It is probably because such feature hides behind our analysis error due to its low spectral density owing to broadly varying H-bonding strengths in liquids.

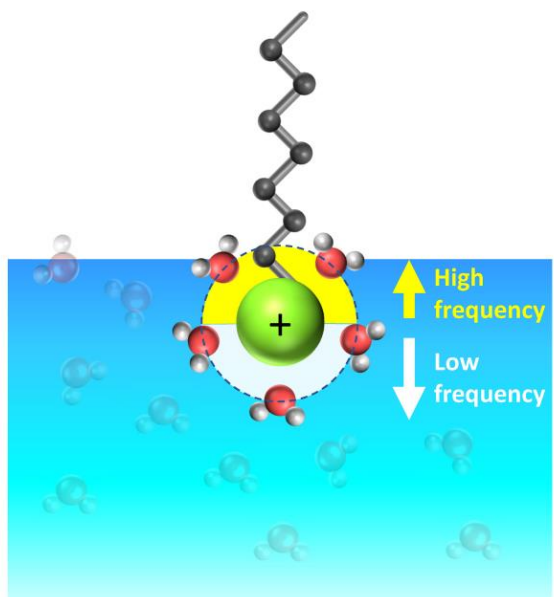

**Figure 6.** Schematic representation of the asymmetric hydration-shell structure of a fully solvated surfactant headgroup, where the up-down symmetry broken by surface-specific bonding environment causes an uncompensated spectral contribution.

A more unified picture could be obtained with helps of the free OH resonance that is found to weaken together with emergence of the 3610–cm$^{-1}$ band in the same bulk concentration range (see Supplementary Section 3 for details). Such correlation is in line with the proposed interpretation suggesting that these two spectral features originate from a distorted local structure at the outmost water molecular layer in terms of creation (annihilation) of hydration water molecules surrounding the CTA$^+$ (unperturbed free OH groups). Through the analysis with discrete-resonance approximation (see Supplementary Section 3), we further deduce a lower-bound estimate of the surface density of water molecules contributing to the 3610–cm$^{-1}$ band, suggesting >13 outmost-layer water molecules affected per single adsorbed CTAB. Note that our scenario is in line with a recent observation on reorientation of free-OH water molecules through the water bending mode.[38]

On the other hand, we also take other headgroup–water bonding configurations into account but fail to find another working explanation for the 3610–cm$^{-1}$ band. First, one may consider a symmetry-broken hydration shell of the Br$^-$ ion. Association of Br$^-$ to the surface CTA$^+$ breaks the up-down symmetry of the hydration shell, resulting in more H-up-oriented water molecules surrounding the Br$^-$. However, the binding interaction of water to the anion was known to cause much lower OH stretching frequencies compared to our observation, as shown, for example, by Saykally and Geissler in a Raman study.[13] Another possibility is water molecules in the second hydration shell surrounding CTA$^+$ with their net upward OH(s) weakly-donor-bonded to the electron-deficient oxygen of water molecules in the first hydration shell, as proposed by Mondal *et. al.* in explaining the effect of metallic ions on bulk water spectra.[41] This scenario relies on strong binding (electron transfer) between the first-shell water molecules and the cation and is, therefore, unlikely in our case due to much weaker binding interactions with the large CTA$^+$ group (5.1 Å in radius[42]).

**Discussion**

To complete the technical development, we give a brief discussion about analysis errors of the MR-SFVS scheme, which depends on $e\phi_0/k_BT$ with $e$ the elementary charge, $k_B$ the Boltzmann constant, and $T$ the temperature. For $e\phi_0/k_BT \ll 1$, the analysis error arises mainly from measurement uncertainty of the reported $\chi_B^{(3)}(\omega_2)$ spectrum.[6] Such uncertainty propagates into the deduced $\phi_0$, but is effectively balanced and thus causes much moderate influences on the retrieved $\chi_{DL}^{(2),A}$ and $\chi_{BIL}^{(2)}$. To get an intuition, one can adopt the Debye-Hückle approximation (valid for $e\phi_0/k_BT \ll 1$) to derive $\Delta\chi_{DL}^{(2)} = \left(\chi_B^{(3)} \cdot \phi_0\right) \cdot [(1 - i\lambda_D\Delta k_z^B)^{-1} - (1 - i\lambda_D\Delta k_z^A)^{-1}]$ and $\chi_{DL}^{(2),A} = \left(\chi_B^{(3)} \cdot \phi_0\right) \cdot (1 - i\lambda_D\Delta k_z^A)^{-1}$ from Eq. (1) and (2) with $\lambda_D$ the Debye screening length. Clearly, when fitting the $\left|\chi_{S,eff}^{(2),B}\right|^2$ spectrum with $\chi_{S,eff}^{(2),A}$, a percentage error of $\chi_B^{(3)}$ is directly transferred to $\phi_0$ via $\Delta\chi_{DL}^{(2)}$ in an inverse manner, such that the errors of $\chi_B^{(3)}$ and $\phi_0$ counteract each other in deducing $\chi_{DL}^{(2),A}$ and $\chi_{BIL}^{(2)}$. As for $e\phi_0/k_BT \gg 1$, the spatial distributions of ions and $E_0(z)$ in DL are known to shrink in depth by following nonlinear PB equations. It results that $\Delta\chi_{DL}^{(2)}$ is no longer proportional to $\phi_0$ but exhibits a sub-linear dependence on $\phi_0$ and, hence, induces appreciable analysis errors in deducing $\phi_0$ and $\chi_{BIL}^{(2)}$ through $\Delta\chi_{DL}^{(2)}$ for high $\phi_0$ (typically > 0.2 V). (See Supplementary Fig. S6, S7 and Ref. [43] for detailed discussions.)

Finally, we comment on recently developed heterodyne second harmonic generation (SHG) spectroscopy that allows deduction of $\phi_0$ of aqueous interfaces.[43-45] Compared to MR-SFVS, nonresonant SHG without specificity to selected resonances would have multiple molecular/material origins,[44,45] e.g., $\chi^{(3)}$ of the solid at a solid/water interface and different molecules in liquid mixtures, making accurate SHG analyses for $\phi_0$ more difficult.

In summary, we have developed a SFVS scheme with varying photon momentums to directly probe vibrational spectra of the bonded interface layer at charged water interfaces. The method does not rely on *prior* information about the interface or assumption of the EDL models, and is viable to any aqueous interfaces (with $e\phi_0$ less than or comparable to $k_BT$). We offered two demonstrations using zwitterionic lipids and model charged surfactants on water and showed that, even in these well-investigated cases, the results provided new insights owing to the spectral separation. For zwitterionic PC monolayers on water, we observed a highly polarized bonded water layer structure associating to the PC headgroup, while the DL contribution was experimentally proven to be negligible. For the CTAB/water interface, our result showed a hidden weakly-donor-H-bonded water species in the BIL, unveiling the hydration structure of the surfactant headgroups. This all-optic technique allows for *in situ* probing of electrochemical and aqueous biological interfaces at the molecular level and opens a new path toward applications with high spatial and ultrafast temporal resolutions.

**Methods**

Our MR-SFVS setup was based on a modified phase-sensitive SFVS with three fundamental inputs prepared by a 1-kHz, 50-fs Ti:sapphire laser system (Astrella, Coherent) equipped with two optical parametric amplifiers (OPAs) and two grating-based spectral filters (GSF). In particular, (1) the femtosecond MIR input pulse with a bandwidth of ~250 cm$^{-1}$ was generated from one OPA pumped by the Ti:sapphire laser, followed by difference frequency generation using $AgGaS_2$. (2) One narrowband fundamental input for SFVS with $l_c^A$ (labeled as "the NIR-A pulse") was generated from the Ti:sapphire output at ~800 nm, with its bandwidth reduced to be ~13.6 cm$^{-1}$ by a GSF. (3) Another narrowband input for SFVS with $l_c^B$ (labeled as "the NIR-B pulse") came from the signal output of the second OPA with a tunable wavelength $\lambda$. Its bandwidth reduced by another GSF was ~30 cm$^{-1}$ that was appropriate to characterize the H-bonded OH-stretching continuum without noticeable spectrum smoothing.

In phase-sensitive SFVS measurements with $l_c^A$ configuration, the MIR and NIR-A pulses were focused onto the sample surface with $\theta = 45°$, after co-propagating through a *y*-cut quartz acting as the local oscillator (LO) and a $SrTiO_3$ plate for phase modulation. The SFG spectral interferogram created by the sample in reflection and the LO reflected from the sample was measured by a charge-coupled device (CCD)-based polychromator. After this data acquisition, a SF intensity measurement of the sample with $l_c^B$ configuration was performed *in situ*, where the condition of the MIR pulse was kept exactly the same, but the NIR-A pulse was replaced by the NIR-B with $\theta = 63.4\ (\pm 1)$ degrees. The SFG from the sample at another emission angle was detected with the same polychromator. The two spectra with $l_c^A$ and $l_c^B$ yielded the complex $\chi_{S,eff}^{(2),A}(\omega_{IR})$ and $\left|\chi_{S,eff}^{(2),B}(\omega_2)\right|^2$ of the sample, respectively, after normalization against a *z*-cut quartz crystal and correction of the Fresnel coefficients. All measurements were performed with S-, S-, P- (SSP-) polarized SFG, near-infrared, and MIR fields at room temperature.

We had $\lambda = 1.3$ μm, $l_c^B = 70.6$ nm, and $l_c^A = 44.5$ nm in water in all of the measurements except a wavelength-dependent test (Fig. 1B,C) where $l_c^B$ was varied in between 65.6 ~ 72.7 nm in water by tuning $\lambda$ (= 1.18~1.37 μm).

**Acknowledgments:** The authors gratefully thank valuable discussions with Dr. Hsien-Ming Lee. **Funding:** This work was funded by National Science and Technology Council, Taiwan (grant number: MOST 108-2112-M-001-039-MY3; 111-2112-M-001-082-). **Author contributions:** Y.H. and Y.C.W. designed the research project and performed the analyses. Y.H., T.H.C., and A.P. conducted the experiments. Y.H., A.P., and Y.C.W. discussed the results, and T.H.C. and Y.C.W. wrote the manuscript. **Competing interests:** The authors declare that they have no competing interest. **Data and materials availability:** All data needed to evaluate the conclusions in the paper are present in the paper and/or the Supplementary Information. Additional data related to this paper may be requested from the authors.

42  Yuan, S. L., Ma, L. X., Zhang, X. Q. & Zheng, L. Q. Molecular dynamics studies on monolayer of cetyltrimethylammonium bromide surfactant formed at the air/water interface. *Colloids and Surfaces a-Physicochemical and Engineering Aspects* **289**, 1-9, doi:10.1016/j.colsurfa.2006.03.055 (2006).

43  Dalstein, L., Chiang, K. Y. & Wen, Y. C. Direct Quantification of Water Surface Charge by Phase-Sensitive Second Harmonic Spectroscopy. *Journal of Physical Chemistry Letters* **10**, 5200-5205, doi:10.1021/acs.jpclett.9b02156 (2019).

44  Ma, E. *et al.* A New Imaginary Term in the Second-Order Nonlinear Susceptibility from Charged Interfaces. *Journal of Physical Chemistry Letters* **12**, 5649-5659, doi:10.1021/acs.jpclett.1c01103 (2021).

45  Wang, H., Hu, X. H. & Wang, H. F. Charge-Induced chi((3)) Susceptibility in Interfacial Nonlinear Optical Spectroscopy Beyond the Bulk Aqueous Contributions: The Case for Silica/Water Interface. *Journal of Physical Chemistry C* **125**, 26208-26215, doi:10.1021/acs.jpcc.1c08263 (2021).

Supplementary Information for

# Momentum-Resolved Sum-Frequency Vibrational Spectroscopy of Bonded Interface Layer at Charged Water Interfaces

Yao Hsiao, Ting-Han Chou, Animesh Patra, and Yu-Chieh Wen[*]

Institute of Physics, Academia Sinica, Taipei 11529, Taiwan, R. O. C.

*Correspondence to: ycwen@phys.sinica.edu.tw (Y.C.W.)

**Supplementary Information for this article includes**

**Section S1. Samples preparation**

**Section S2. Interpretation of the BIL spectrum of the HePC monolayer on water**

**Section S3. SF spectrum analysis of CTAB solutions**

**Additional figures**

**Section S1. Samples preparation**

Lignoceric acid (LA, > 99% purity), cetyltrimethylammonium bromide (CTAB, > 99%), hexadecyl phosphocholine (HePC, > 99%), HCl (37 wt % water solution, reagent grade), NaOH (reagent grade, pellets), and chloroform (anhydrous grade, stabilized with ethanol) were obtained from Sigma-Aldrich. 1,2-dipalmitoyl-*sn*-glycero-3-phosphocholine (DPPC) was purchased as lyophilized powders from Avanti Polar Lipids. All chemicals were used as received. Water was deionized by a Millipore system and had a resistivity of 18.2 MΩ-cm. Its pH was varied by solvation of NaOH and HCl.

We follow Ref. [1] and [2] to prepare Langmuir monolayers on water in a thoroughly cleaned Teflon Langmuir trough. LA or DPPC was spread from a chloroform solution onto the aqueous surface, and the solvent was then allowed to evaporate for ~10 minutes. The monolayer of LA (DPPC) was then compressed to the surface pressure of ~27 (~28) mN/m, corresponding to a closely packed monolayer in the tilted condensed phase. The CTAB and HePC monolayers on water were formed by following the Gibbs adsorption kinetics with a surface coverage controlled by the bulk concentration.

**Section S2. Interpretation of the BIL spectrum of the HePC monolayer on water**

Based on $\chi^{(2)}_{S,eff}(\omega_2) \cong \chi^{(2)}_{BIL}(\omega_2)$ confirmed in the main text for the PC monolayers on water, we give a brief discussion about the BIL structure through the $\text{Im}\chi^{(2)}_{S,eff}(\omega_2)$ spectrum. We follow the spectrum interpretation of the PC lipid monolayer/water interface in Ref. [3-5] to examine our result of the HePC/water interface for its structural simplicity. Shown in Fig. S1 is the measured $\text{Im}\chi^{(2)}_{BIL}(\omega_2)$ spectrum of the HePC/water interface with a bulk HePC concentration of 1.6 μM (about 12 % of the critical micelle concentration). It displays a dominant positive band at 3000 ~ 3450 cm$^{-1}$, a weak negative band at 3450 ~ 3700 cm$^{-1}$, and a small sharp peak at ~3700 cm$^{-1}$. The peak must arise from residual dangling OHs of the outmost-layer water molecules for our sample with a sub-monolayer surface coverage. The weak negative band at 3450 ~ 3700 cm$^{-1}$ are attributed to downward-pointing OHs (toward bulk liquid) of water molecules acceptor-bonded to choline.[4,5] Interestingly, it is found that the positive $\text{Im}\chi^{(2)}_{BIL}$ in the 3000 – 3450 cm$^{-1}$ region is very intense (5.6×10$^{-21}$ m$^2$/V). It is readily comparable with the SF responses of many charged aqueous interfaces with significant $\chi^{(3)}_B$ contributions, such as fully deprotonated fatty acid monolayer on 10-mM NaOH solution and the surface of a 0.1-mM CTAB solution, as also shown in Fig. S1 for comparison. (Note that the bulk NaOH or CTAB concentrations chosen in these examples correspond to the optimized conditions for the maximum SF OH-stretching responses.) Strong $\text{Im}\chi^{(2)}_{BIL}$ of the HePC/water interface reflects giant polarization from H-bonded water molecules with their O→H pointing upward. The similar feature was also observed in the PC lipids/water interface.[3-5] ($\text{Im}\chi^{(2)}_{BIL}$ of the DPPC/water interface at 3200 cm$^{-1}$ is about half of that of the HePC/water interface, not shown here.) Recently, the water molecules contributing to this band were found to exhibit an up-down reversal of their net orientation by following the chemical inversion of the charged phosphate and choline in the headgroup,[3] suggesting that their orientation is dictated by the local dipolar field generated between the two charged groups and/or a H-bonding structure bridging both phosphate and choline, as also pointed out by Morita *et. al.* in a molecular dynamics calculation.[4]

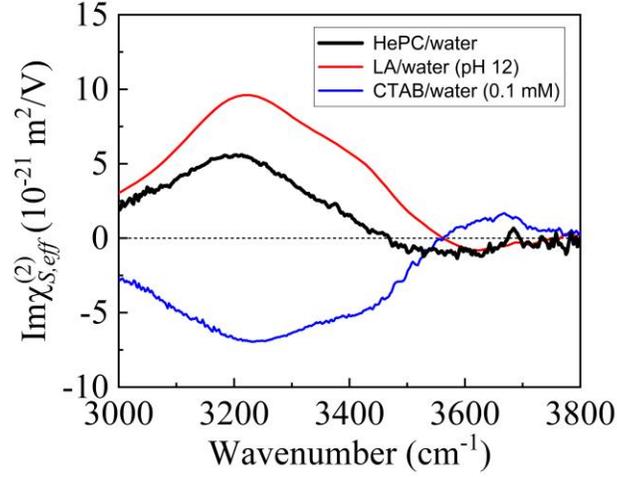

**Figure S1.** Measured $\text{Im}\chi^{(2)}_{S,eff}(\omega_2)$ spectrum of the surface of a 1.6 μM HePC solution at pH 7.8, in comparison with the lignoceric acid (LA)/water interface at bulk pH 12 and the surface of a 0.1-mM CTAB solution.

**Section S3. SF spectrum analysis of CTAB solutions**

To characterize the free OH resonance and the new 3610–cm$^{-1}$ band for the surface of CTAB solutions, we follow Ref. [6] to fit the $\chi^{(2)}_{BIL}(\omega_{IR})$ spectra by approximating the OH-stretching resonances as discrete Lorentzian functions:

$$\chi^{(2)}_{BIL}(\omega_{IR}) = \chi^{(2)}_{NR} + \sum_q \frac{A_q}{\omega_{IR}-\omega_q+i\Gamma_q},$$

(S1)

where $\chi^{(2)}_{NR}$ is the nonresonant background, $A_q$ is the mode amplitude, $\omega_q$ is the resonant frequency, and $\Gamma_q$ is the damping coefficient. We first fit the spectrum of the neat water/air interface as a reference using four Lorentzian functions ($q$ = 1~4, with $\omega_1 = 3315$ cm$^{-1}$, $\omega_2 = 3455$ cm$^{-1}$, $\omega_3 = 3613$ cm$^{-1}$, and $\omega_4 = 3704$ cm$^{-1}$. Note that the subscript refers to the resonance modes in this Supplementary Section.) For the CTAB/water interface, we approximate the mode parameters ($A_q$, $\omega_q$, and $\Gamma_q$) for $q = 1\sim3$ remain the same as the neat water surface, but consider that the CTAB adsorption can alter the mode parameters for the free OH resonance ($q = 4$) and induce a new band ($q = 5$) at ~3610 cm$^{-1}$. Shown in Fig. S2 are the deduced mode parameters for $q = 4$ and 5. It is seen that the 3610–cm$^{-1}$ band emerges above our detection limit for [CTAB] > 3 μM, and in a similar concentration range, the free OH resonance starts to exhibit discernible spectral changes in terms of $A_4$ and $\omega_4$. Particularly, the free OH resonance red-shifts by 4±2 cm$^{-1}$ with its amplitude reduced by 16±6 % as increasing the CTAB concentration from 0 to 30 μM.

Knowing that the mode amplitude is associated with the surface density of OHs and ensemble orientational average of the molecular hyperpolarizability,[6] we can use the mode amplitudes deduced from the fitting to estimate the *lower bound* of the surface density of OHs contributing to the new 3610-cm$^{-1}$ band for the CTAB/water interface. We adopt the mode amplitude of the free OH resonance at the neat water/air interface, labelled as $A_4^0$, as the reference and compare it with the mode amplitude of the 3610–cm$^{-1}$ band, $A_5$, for the CTAB solutions. Approximately, we ignore frequency dependency of the molecular hyperpolarizability because of the moderate frequency difference ($\omega_4 - \omega_5 \approx 100$ cm$^{-1}$) and consider vertically aligned dangling OH groups without much calculation error.[6] The ratio $A_5/A_4^0$ then reflects the lower bound of the ratio of the surface densities of OHs contributing to the two corresponding resonances. (It is a lower-bound estimate due to the orientational average of OHs for the 3610–cm$^{-1}$ band.) With the known densities of free OHs (2.5 nm$^{-2}$) and water molecules (10 nm$^{-2}$) in the outmost layer of the neat water surface, the surface density of OHs contributing to the 3610–cm$^{-1}$ band for the CTAB/water interface is estimated to be > 6.3 nm$^{-2}$ for [CTAB] = 30 μM. This estimate corresponds to >13 water molecules affected per single adsorbed CTAB molecule as taking the CTAB surface excess of ~0.24 nm$^{-2}$ into account (estimated by surface tension through the Gibbs adsorption isotherm[7]).

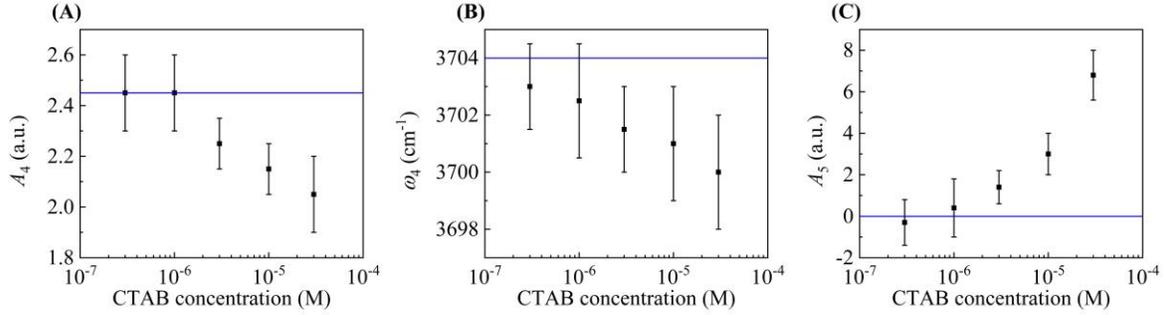

**Figure S2.** Mode parameters (*A*) $A_4$, (*B*) $\omega_4$, and (*C*) $A_5$ deduced by fitting the $\chi^{(2)}_{BIL}(\omega_{IR})$ spectra of the CTAB/water interface for different CTAB concentrations in water (dot). $\Gamma_4 = 18$ cm$^{-1}$, $\Gamma_5 = 120$ cm$^{-1}$, and $\omega_5 = 3610$ cm$^{-1}$ are the same throughout the fitting. Lines depict the values for the neat water/air interface. Note that $q = 4$ and 5 refer to the free OH resonance and the new 3610–cm$^{-1}$ band, respectively.

**Additional figures**

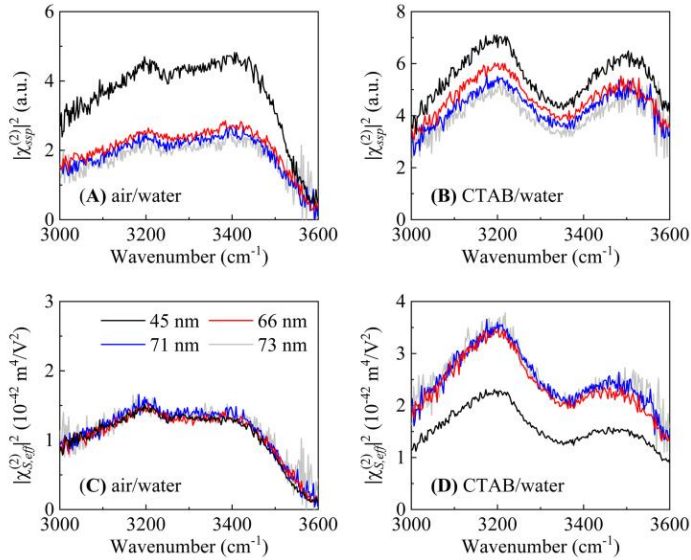

**Figure S3.** SF intensity spectra of (*A*), (*C*) the neat water/air interface and (*B*), (*D*) the surface of a 10-μM CTAB solution measured by MR-SFVS with different coherent lengths. (*A*) and (*B*) The raw SF intensity

after normalization against the quartz reference. (*C*) and (*D*) The results deduced from *A* and *B* after correcting the Fresnel coefficients, and the nonlinear optical susceptibility and the coherence length in quartz.

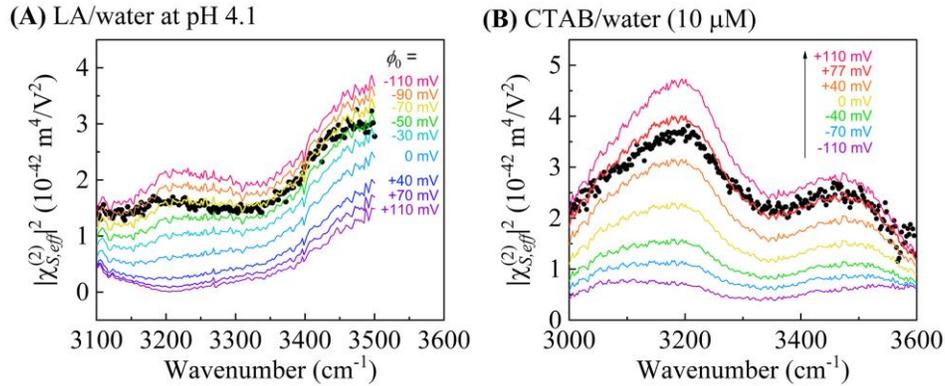

**Figure S4.** $\left|\chi_{S,eff}^{(2),B}(\omega_2)\right|^2$ SF intensity spectra of (*A*) the LA/water interface at pH 4.1 and (*B*) the surface of a 10-μM CTAB solution measured with $l_c = l_c^B$ (dots), in comparison with spectral fits calculated from Eq. (2) in the main text with the measured $\chi_{S,eff}^{(2),A}(\omega_2)$ and the known[1] $\chi_B^{(3)}(\omega_2)$ for different values of the diffuse layer potential $\phi_0$ (lines). Fitting the measured $\left|\chi_{S,eff}^{(2),B}(\omega_2)\right|^2$ allows for determining both sign and magnitude of the only fitting parameter $\phi_0$. Negative- (positive-)valued $\phi_0$ deduced from *A* (*B*) corresponds to a negatively- (positively-)charged LA/water (CTAB/water) interface, as expected. Note that the fits calculated with $\phi_0 = 0$ V represent the measured $\left|\chi_{S,eff}^{(2),A}(\omega_2)\right|^2$ spectra.

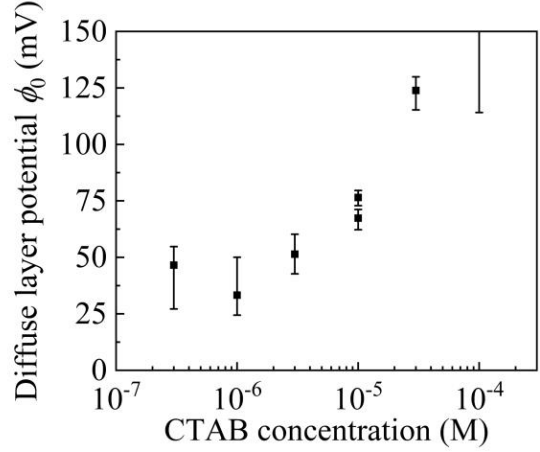

**Figure S5.** Diffuse layer potential $\phi_0$ at the air/water interface for different CTAB concentrations in water deduced by the MR-SFVS analysis. Detailed discussions about $\phi_0$ rely on the EDL models for Gibbs monolayers on water that may take thick adsorption layer and/or ion binding into account.[8] It is beyond the scope of this paper.

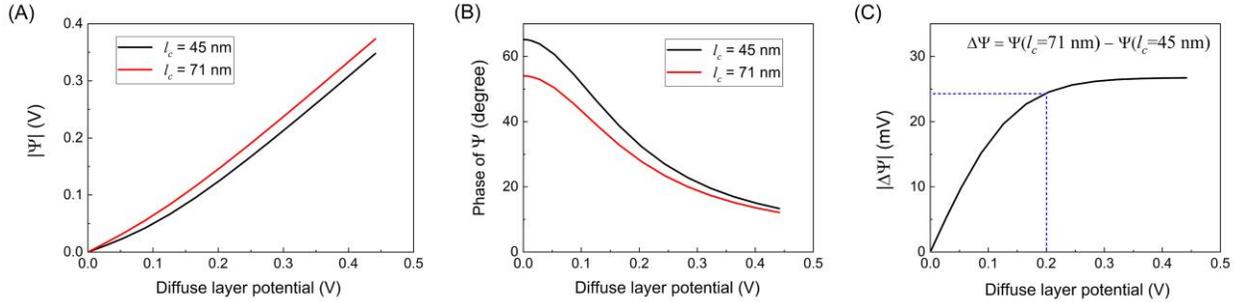

**Figure S6.** (*A*) Amplitude and (*B*) phase of $\Psi$ versus the diffuse layer potential $\phi_0$ calculated from Eq. (1) in the main text using the PB theory with an ionic strength of 10 μM for two different coherence lengths. (*C*) Amplitude of the differential $\Psi$, $\Delta\Psi \equiv \Psi(l_c = 71\text{nm}) - \Psi(l_c = 45\text{nm})$, deduced from *A* and *B*. The phase of $\Psi$ is seen to decrease with an increase in $\phi_0$, reflecting a reduced thickness of the DL for higher $\phi_0$ due to the nonlinear PB equation. This phenomenon results in a sublinear (saturation) behavior of $\Delta\Psi$ and $\Delta\chi_{DL}^{(2)}$ ($\propto \Delta\Psi$) with respect to $\phi_0$ for high $\phi_0$, as seen in *C* for $\phi_0 > \sim 0.2$ V.

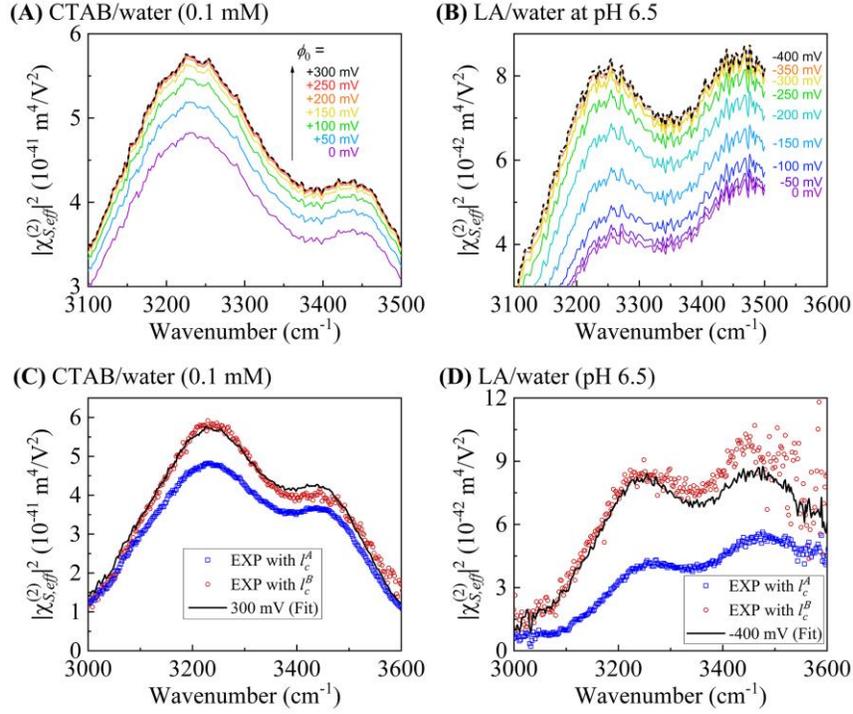

**Figure S7.** (*A*) and (*B*) Calculated $\left|\chi^{(2),B}_{S,eff}(\omega_2)\right|^2$ spectra of (*A*) the surface of a 0.1-mM CTAB solution and (*B*) the LA/water interface at pH 6.5 for different $\phi_0$ (solid lines). The calculation is based on Eq. (2) in the main text with the measured $\chi^{(2),A}_{S,eff}(\omega_2)$ and the known[1] $\chi^{(3)}_B(\omega_2)$. The results are seen to follow the sub-linear dependence of $\Delta\Psi$ on $\phi_0$, as discussed in Fig. S6, to vary with respect to $\phi_0$ for low $\phi_0$ but become saturated for high $\phi_0$. Dashed lines denote the calculation results in the saturation regime (with $\phi_0$ chosen somehow arbitrarily). (*C*) and (*D*) Measured SF intensity spectra of the two samples for $l_c = l_c^A$ and $l_c^B$ (dots), in comparison with the calculated $\left|\chi^{(2),B}_{S,eff}(\omega_2)\right|^2$ from *A* and *B* in the saturation regime (lines). Quantitative consistency between the measured $\left|\chi^{(2),B}_{S,eff}(\omega_2)\right|^2$ and the calculation in the saturation regime is found for the two samples here, and so for other high-$\phi_0$ CTAB solutions with [CTAB] = 0.1~1 mM (not shown here). This agreement confirms the theoretical understanding and validity of the $\chi^{(3)}_B(\omega_2)$ spectrum.